# Trajectories of loose sand samples in the Phase Space of Soil Mechanics :

## P. Evesque

Lab MSSMat, UMR 8579 CNRS, Ecole Centrale Paris
92295 CHATENAY-MALABRY, France, e-mail evesque@mssmat.ecp.fr

**Abstract:**
*In general, the evolution of soil submitted to simple stress-strain paths is characterised using the 3d phase space (v,p',q) i.e. (specific volume, mean intergranular pressure, deviatoric stress $q=\sigma'_1-\sigma'_2$. When compressions is performed at $\sigma'_2=\sigma'_3=c^{ste}$ or at $p'=(\sigma'_1+\sigma'_2+\sigma'_3)/3=c^{ste}$, one finds that all trajectories end up at a line of attracting point called the critical-state line via the surface of Roscoe or of Hvorslev depending if the initial volume is the loosest possible one (at a given p') or denser. Trajectories of weakly dense samples is not often reported in this phase space. We find here that it shall present some sigmoid shape as it can be found from soil mechanics argument. This seems to indicate that Roscoe's surface shall exhibit a singularity at the critical point.*
______________________________________________________________________

Triaxial apparatus is often used to characterise the mechanical behaviour of granular media within axisymmetric compression [1-3]. Experiments performed at $\sigma'_2=\sigma'_3=c^{ste}$ or at $p'=(\sigma'_1+\sigma'_2+\sigma'_3)/3$ ends up at the critical state which is defined by the specific volume [1-4]:

$$v_c=v_{co}-\lambda \ln(p'/p'_o) = v_{nco}-\lambda \ln(p'/p'_o)-\lambda_d \ln(2) \qquad (1)$$

where $v_{co}$, $v_{nco}$, $\lambda$ and $\lambda_d$ are constancies which depend on the material. One defines also the "normally consolidated state" as the loosest possible state under an isotropic loading p'. Its specific volume $v_{nc}$ is given by:

$$v_{nc}= v_{nco}-\lambda \ln(p'/p'_o) \qquad (2)$$

$\lambda_d$ is always positive so that the specific volume of the loosest possible samples decreases always under a deviatoric compression, i.e. $q=\sigma'_1-\sigma'_2$, according to the law:

$$v= v_{co}-\lambda \ln(p'/p'_o)-\lambda_d \ln[1+q^2/(p'^2M'^2)] \qquad (3)$$

where M' is related to the friction angle $\varphi$ by $M'=6\sin\varphi/(3-\sin\varphi)$. An other important relation which seems to be satisfied is the Rowe's relation, which states that during a compression test performed at $\sigma'_2=\sigma'_3=c^{ste}$, the dilatancy, defined as $-\partial\varepsilon_v/\partial\varepsilon_1$, depends only on the q/p' ratio, since it follows the law:

$$1-\partial\varepsilon_v/\partial\varepsilon_1=\tan^2(\pi/4-\varphi/2)\,(\sigma'_1)/\sigma'_2 \qquad (4)$$

where $\varepsilon_1$ ($\varepsilon_v$) is the axial (volumic) deformation. According to this, one concludes [1-3] in general that the sample shall pass through a minimum of volume and pass through also a maximum of deviatoric stress q before reaching the plasticity plateau defined as q/p'=M' and v=$v_c$ as soon as the sample is denser than the loosest possible sample. The minimum volume and the deviatoric-stress maximum are not concomitant. In general also, one reports the data in the specific volume v, mean pressure p' plane .

Considering then a $\sigma'_2=\sigma'_3=c^{ste}$ compression test, and starting from a relatively loose sample whose specific volume $v_i$ ranges in between $v_{nc}(p')>v_i>v_c(p')$, this lets predict a sigmoid shape for the projection of the trajectory in this {v,ln(p')} plane. Indeed, starting from a mean pressure $\sigma'_2=\sigma'_3=p'_o$ , one shall have all along this test:

$p'=p'_o+q/3$ (5.a)

$q=3(p'-p'_o)$ (5.b)

$\partial\varepsilon_v/\partial\varepsilon_1=1- \tan^2(\pi/4-\varphi/2)\ (3p'+2p'_o)/p'_o$ (5.c)

and the trajectory shall stop at:

$q_c=3p'_o/(3-M')$ (5.d)

$p'_c=p'_o(4-M')/(3-M')$ (5.e)

$v_c=v_{nc}(p'_o)+ \lambda\ [\ln(3-M')-\ln(4-M')]-\lambda_d \ln(2)$ (5.f)

So, writing $d\varepsilon_v= (\partial\varepsilon_v/\partial\varepsilon_1)\ (\partial\varepsilon_1/\partial q)\ (\partial q/\partial p)\ dp$ , one gets:

$d\varepsilon_v =3dp'\ [1- \tan^2(\pi/4-\varphi/2)\ (3p'+2p'_o)/p'_o]\ (\partial\varepsilon_1/\partial q)$ (6)

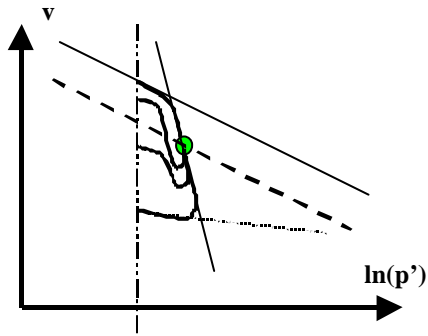

**Figure 1:** *Predicted trajectories in the {v,ln(p')} plane for a compression test at $\sigma'_2=\sigma'_3=c^{ste}$, for different samples of the same sand packed at different initial densities, starting from the same $\sigma'_2=p'_o$ . The inclinecd full (dashed) line represents the normally consolidated (critical) states. Bold full curves are the trajectories. For loose samples, but denser than the loosest one, the trajectory shall start with a slope smaller than the normally consolidated slope; they shall exhibit a minimum of volume $v_{min}<v_c$ and a maximum of p' such as $p'_{max}>p'_c$ ; they shall end at $p'=p'_c$ and $v=v_c$ . So, they shall exhibit a sigmoid shape.*

The term in between the bracket [ ] changes of sign at the characteristic state, i.e. when q/p'=M' . The last term in the right member becomes infinite at the peak of deviatoric stress, which corresponds also to the maximum of p' according to Eq. (5.a), and at the critical state. The trajectory has a horizontal slope at the minimum of volume and a vertical slope at the maximum of p'. It could have a vertical slope at the critical state; however, it has not since the term in bracket [ ] is equal to 0 in that peculiar case. This leads to the trajectories reported in Fig. 1. It is worth mentioning also that the minimum of volume occurs when $p'=p'_c$ owing to the Rowe's relation.

So, accordingly to soil mechanics, one expects that the trajectories shall start with a slope less inclined than the one of the normally-consolidated-state line since the sample looks harder (due to over-consolidation). They shall end at the critical state, after having passed through a maximum of density $v_{min}<v_c$, occuring at $p'_c$, and through a maximum of pressure $p'_{max}>p'_c$ slightly after the minimum of volume. So, it seems that the sigmoid shape is needed for large initial v, i.e. $v_c<v_{initial}<v_{nc}$.

They are not so many trajectories published in the literature, which correspond to such behaviours, although loose samples have been investigated. It is probably due to the fact that it is difficult to measure without uncertainty the initial density of a loose sample and to measure also the critical density with enough accuracy to get absolute and reliable data. One can find such data in Saïm thesis [5], for instance on Figs. 5.24 to 5.26, 5.31, 5.32 and 5.35 to 5.38 .

Anyhow, it is not completely obvious that the first part of the trajectory shall be as flat as this paper predicts near q=0; it can be more inclined, perhaps. Saïm [5] finds it rather horizontal in agreement with the present prediction however. But this depends obviously on the real evolution of $(\partial\epsilon_1/\partial q)$ with q.

An other point which is rather surprising is the difference of shape of the predicted trajectories for the "normally consolidated state" and the ones which are slightly denser: in the first case, no sigmoid shape is predicted and arrival at the critical state occurs from the top, and not from the bottom. This is indeed what is observed in the experimental data reported by Saïm [5]. It seems then that one should be able to define a specific initial density $v_b$ at which a bifurcation occurs: for initial specific volume $v_i$ such as $v_i>$ (or <) $v_b$ the trajectories shall arrive at the critical state from the top (or from the bottom). If $v_b < v_{nc}$, it would mean that Rowe's relation is not strictly valid. On the contrary, if the Rowe's relation remains valid, this would imply that $v_b=v_{nc}$.

In such a case, *i.e.* $v_b=v_{nc}$, it exists an other difficulty: why does $v=v_{nc}$ is the only initial density which can reach the critical state from the top?? Indeed it is more reasonable to consider in that case that **all trajectories** corresponding to $\sigma'_2=\sigma'_3=p'_o=c^{ste}$ compression **arrives at the critical state from the bottom** in the {v,ln(p')} plane and that the **normally consolidated** sample is just the limit case when the minimum of volume corresponds also to the critical volume.

If this last analysis is true, it will have some consequence. In paticular, it may explain the difficulty of getting critical flow at a free surface and explain the systematic existence of avalanches: stable slopes at the critical angle can only be made statically stable by passing through a denser density, generating always a dilatancy mechanism for the next avalanche.

As a conclusion, this paper tries and emphasises that loose sand samples shall exhibit some remarkable features in the {v,ln(p')} plane, with a sigmoid shape. If this sigmoid shape is not obtained for very loose samples, it may indicate that the Rowe's relation does not hold exactly; on the contrary, if the sigmoid shape is already obtained for very lose samples, it will indicate that the normally consolidated behaviour exhibits a very peculiar behaviour, which has been never considered previously but which as some important consequences from the view point of the mechanics. In this last case in particular, this behaviour would probably be able to explain the spontaneous generation of avalanching process at the free surface of an inclined pile and the hysteretical behaviour of the flow.

More experimental data, with better accuracy, are then needed to raise this uncertainty.

*Acknowledgements:* CNES is thanked for partial funding.

The electronic arXiv.org version of this paper has been settled during a stay at the Kavli Institute of Theoretical Physics of the University of California at Santa Barbara (KITP-UCSB), in june 2005, supported in part by the National Science Fundation under Grant n° PHY99-07949.

*Poudres & Grains* can be found at :
http://www.mssmat.ecp.fr/rubrique.php3?id_rubrique=402